\documentclass{PoS}
\usepackage{bm}
\usepackage{amsmath,amsfonts,amssymb,amscd,amsxtra,amsthm}

\title{On a possibility of exotic heavy baryons}

\ShortTitle{Exotic heavy baryons}

\author{\speaker{Michal Praszalowicz}\\ 
        Marian Smoluchowski Institute of Physics, Jagiellonian
  University, \\{\L}ojasiewicza 11, 30-348 Krak{\'o}w, Poland\\
        E-mail: \email{michal@if.uj.edd.pl}}

\abstract{Triggered by a recent announcement by the LHCb Collaboration of five excited $\Omega^0_c$ resonances of  small widths
we propose an interpretation based on the Chiral Quark Soliton Model. We argue that three of the LHCb
resonances are parity ($-$) excitations of the ground state SU(3) sextets, while the other two with the smallest
widths are exotic pentaquarks that belong to the SU(3) $\overline{\mathbf{15}}$. We first briefly review
the model and discuss the status of the putative $\Theta^+(1540)$. Next we show how to generalize the
model to the case of heavy baryons. We test this approach against the ground state  $\overline{\mathbf{3}}$ and ${\mathbf{6}}$
heavy baryons, and discuss different excitations that are possible in the Chiral Quark Soliton Model.
We show that the model accommodates two excited  $\overline{\mathbf{3}}$ in the charm sector that are experimentally
observed. Finally
we discuss possible assignments of the LHCb $\Omega^0_c$ states.}

\FullConference{Corfu Summer Institute 2017 'School and Workshops on Elementary Particle Physics and Gravity'\\
		2-28 September 2017\\
		Corfu, Greece}

\begin{document}

\section{Introduction}
In a recent paper the LHCb collaboration announced five, or even six
$\Omega^{0}_{c}$ states with masses in the range of $3 - 3.2$~GeV
\cite{Aaij:2017nav}. These states have been later confirmed by BELLE 
\cite{Yelton:2017qxg}. In this report we summarize our recent works
on heavy baryons \cite{Yang:2016qdz,Kim:2017jpx,Kim:2017khv}
(see also \cite{Kim:2018xlc,Yang:2018uoj,Kim:2018nqf})
where we have applied the Chiral Quark Soliton Model ($\chi$QSM)
to the baryonic systems with one heavy quark. This paper is based
on Ref.~\cite{Praszalowicz:2017cwk}, a shortened version of this
report has already appeared \cite{Praszalowicz:2018azt}.

The LHCb discovery triggered theoretical activity to interpret  $\Omega^0_c$'s  in different 
approaches. These include  QCD sum rules~\cite{Wang:2017zjw,
    Chen:2017sci, Agaev:2017jyt,Aliev:2017led,Mao:2017wbz}, constituent
quark models~\cite{Wang:2017hej}, chiral quark models \cite{Yang:2017xpp},  lattice QCD
\cite{Padmanath:2017lng} and holographic QCD \cite{Liu:2017xzo}.
In Refs.~\cite{Karliner:2017kfm,Wang:2017vnc,Cheng:2017ove,Ali:2017wsf} 
the new states are treated as bound states of
a charm quark and a light diquark, the authors of
Refs.~\cite{Huang:2017dwn,Montana:2017kjw,Wang:2017smo,Chen:2017xat} 
interpreted the new states as 
molecular states and in some
approaches~\cite{Yang:2017rpg,An:2017lwg,Anisovich:2017aqa,Wang:2018alb} as pentaquarks. 
Regge trajectories approach has been put forward in Ref.~\cite{Chen:2017fcs}.
Decay properties
of these states have been studied in a $^3P_0$ model \cite{Zhao:2017fov},
quark potential models \cite{Chen:2017gnu}, chiral quark model \cite{Wang:2017kfr}
and in molecular model \cite{Huang:2018wgr}.
Coupled channel approach to study the LHCb $\Omega^0_c$ states has been applied in 
Refs.~\cite{Debastiani:2017ewu,Nieves:2017jjx}.
For theoretical  review
of heavy exotica see Ref.~\cite{Ali:2017jda} and for experimental overview Ref.~\cite{Olsen:2017bmm}.

Such a large number of possible theoretical interpretations may at first glance come as a surprise.
Indeed,  naive quark model seems to provide the simplest and successful picture. Let us recall
that two light quarks may form SU(3)$_{\rm flavor}$ $\overline{\mathbf{3}}$ and ${\mathbf{6}}$.
Because of the Fermi exclusion principle $\overline{\mathbf{3}}$ has spin 0 and ${\mathbf{6}}$
spin 1. Adding a heavy quark results in one SU(3) antitriplet of spin 1/2 and two hyperfine split
sextets of spin 1/2 and 3/2. The simplest possible excitation consists in adding the angular momentum 
to the system, which in the heavy quark rest frame  maybe interpreted as the angular momentum of
the light subsystem. Such configuration would have negative parity. An immediate consequence  
of this picture  is the emergence of two hyper-fine
split antitriplets of spin $1/2^-$  and $3/2^-$ that are indeed observed experimentally both in charm
and (partially) bottom sectors. In the sextet case total angular momentum of the light subsystem can be 0, 1 or 2.
Therefore one predicts five excited sextets of negative parity: two with total spin $1/2$, two with total
spin 3/2 and one with total spin 5/2. 
The LHCb resonances would be the first experimentally observed particles from these multiplets.
Unfortunately, when it comes to a more detailed  analysis of the LHCb
data, basically all models have problems to accommodate all five LHCb resonances within the above scenario
with acceptable accuracy. Therefore an alternative assignments of some of the LHCb resonaces have been necessary.

 We have encountered similar problems in the $\chi$QSM. In a recent
paper \cite{Yang:2016qdz} we have shown that   the ground state sextets
together with the ground state $\overline{\mathbf{3}}$ that comprises  
$\Lambda_{c}(2280)$ and $\Xi_{c}(2470)$ can be successfully described in terms
of the $\chi$QSM
supplemented by an interaction with a heavy quark in such a way
that heavy quark symmetry  \cite{Isgur:1989vq} is respected.
 A great advantage
of the $\chi$QSM consists in a rather restrictive mass formula linking
the spectra of light baryons with the heavy ones in question, which 
-- as mentioned above -- does
not allow to accommodate all five LHCb $\Omega^0_c$ resonances in
the parity $(-)$ excitations of the ground state sextet. Therefore we have been
forced to find an alternative assignment for some of these states. A natural
interpretation within the framework of the $\chi$QSM is to associate two narrowest
(with the decay width of the order of 1 MeV) $\Omega^0_c$'s with parity $(+)$ exotic
pentaquarks.

\section{Chiral Quark Soliton Model for light baryons}

$\chi$QSM \cite{Diakonov:1987ty} (for  review see Refs.~\cite{Christov:1995vm,Alkofer:1994ph,Petrov:2016vvl} and references therein)
 is based on an old argument by Witten~\cite{Witten:1979kh}, which says that in the limit of a large number 
 of colors ($N_c \rightarrow \infty$), 
 $N_c$ relativistic valence quarks generate chiral mean fields represented by a distortion of 
 the Dirac sea that in turn interacts with the valence quarks themselves.
The soliton configuration corresponds to the solution of the Dirac equation for the constituent quarks (with
gluons integrated out) in the mean-field approximation where the mean fields respect so called {\em hedgehog}
symmetry. Since it is impossible to construct a pseudoscalar field that changes sign under
inversion of coordinates, which would be compatible with
the SU(3)$_{\rm flav}\times$SO(3) space symmetry one has to resort to a smaller {\em hedgehog}  symmetry that, however,
leads to the correct baryon spectrum (see below).
This means that 
neither spin ($\bm{S}$) nor isospin ($\bm{T}$)  are  {\em good} quantum numbers. Instead a {\em grand spin}  
$\bm{K}=\bm{S}+\bm{T}$ is a {\em good} quantum number.

\begin{figure}[h]
\centering
\includegraphics[width=9.0cm]{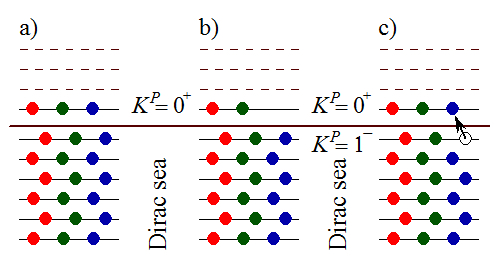} \vspace{-0.2cm}\caption{Schematic
pattern of light  quark levels in a self-consistent soliton
configuration. In the left panel all sea levels are filled and $N_{c}$ (=3 in
the Figure) valence quarks occupy the $K^{P}=0^{+}$ lowest
positive energy level. Unoccupied positive energy levels are dpicted
by dashed lines. In the middle panel one valence quark has been
stripped off, and the soliton has to be 
supplemented by a heavy quark not shown in the Figure. In the right panel a
possible excitation of a sea level quark, conjectured to be $K^{P}=1^{-}$, to
the valence level is shown, and again the soliton has to couple to a heavy
quark. Strange quark levels that exhibit different filling pattern are
not shown.}%
\label{fig:levels}%
\end{figure}

The ground state configuration corresponds to the fully occupied $K^P=0^+$ valence level, as shown in Fig.~\ref{fig:levels}.a.
Therefore the soliton does not carry definite quantum numbers except for the baryon number resulting from the valence quarks.
Spin and isospin appear when the rotations in space and flavor are quantized 
and the resulting {\em collective} hamiltonian analogous to the one
of a symmetric top is computed. There are two conditions that the {\em collective} wave functions
have to satisfy:
\begin{itemize}
\item allowed SU(3) representations must contain states with hypercharge
$Y^{\prime}=N_{c}/3$,
\item the isospin $\bm{T}^{\prime}$ of the states with $Y^{\prime}%
=N_{c}/3$ couples with the soliton spin $\bm{J}$ to a singlet:
$\bm{T}^{\prime}+\bm{J}=0$.
\end{itemize}
\begin{figure}[h!]
\centering
\includegraphics[width=10cm]{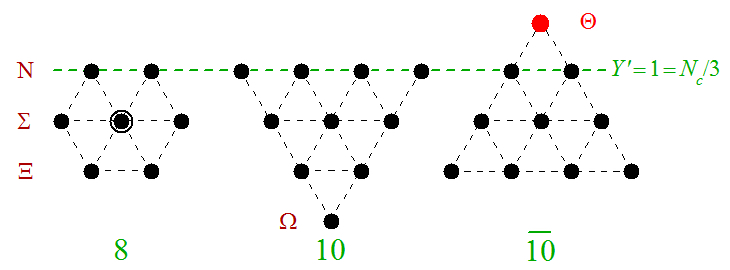} \vspace{-0.1cm}
\caption{Lowest lying SU(3) flavor representations allowed by the constraint $Y'=1$. The first 
{\em exotic} representation, $\overline{\mathbf{10}}$, contains the explicitly
exotic pentaquark states ${\Theta}^+$,
${\Xi}^{+}$  and ${\Xi}^{-}$ and non-exotic nucleon- and sigma-like states.}
\label{fig:irreps1}%
\end{figure}

In the case of light positive parity baryons the lowest allowed representations are
$\mathbf{8}$ of spin 1/2, $\mathbf{10}$ of spin 3/2, and also exotic
$\overline{\mathbf{10}}$ of spin 1/2 with the lightest state corresponding to
the putative ${\Theta}^{+}(1540)$~\cite{Praszalowicz:2003ik,Diakonov:1997mm}. 
They are shown in Fig.~\ref{fig:irreps1}. 
Chiral models in general predict that pentaquarks are light 
\cite{Praszalowicz:2003ik,Diakonov:1997mm}
and -- in some specific models -- narrow \cite{Diakonov:1997mm}.

After 
the first enthusiastic announcements of the discovery of pentaquarks in 2003 
by LEPS
\cite{Nakano:2003qx} and DIANA \cite{Barmin:2003vv} collaborations,
the experimental evidence for the light
exotica has been questioned (see {\em e.g.} \cite{Hicks:2012zz}). Nevertheless,
both DIANA \cite{Barmin:2013lva} and LEPS \cite{Nakano:2017fui} upheld their 
original claims after performing higher statistics analyses. The report on
exotic ${\Xi}$ states (see Fig.~\ref{fig:irreps1}) by NA49 \cite{Alt:2003vb}
from 2004, to the best of my knowledge, has not been questioned so far, 
however the confirmation is still strongly needed.

Another piece of information on $\overline{\mathbf{10}}$ comes from the
$\eta$ photo-production off the nucleon.  Different experiments confirm the narrow
structure at the c.m.s. energy $W \sim  1.68$~GeV
observed in the case of the neutron, whereas no structure is observed on the
proton (see Fig.~27 in the latest report by CBELSA/TAPS Collaboration \cite{Witthauer:2017pcy} 
and  references therein). The natural interpretation of this "neutron puzzle" was proposed already
in 2003 in Ref.~\cite{Polyakov:2003dx}.
There one assumes that the narrow excitation at $W \sim  1.68$~GeV corresponds to the non-exotic
penta-nucleon resonance belonging to $\overline{\mathbf{10}}$.
Indeed, the SU(3) symmetry forbids photo-excitation of the proton member of $\overline{\mathbf{10}}$, while
the analogous transition on the neutron is possible. This is due to the fact that photon is an SU(3)  $U$-spin singlet,
and the $U$-spin symmetry is exact in the SU(3) symmetric limit. An alternative interpretation is based
on a partial wave analysis in terms of  the Bonn-Gatchina approach \cite{Anisovich:2017xqg}. 
There is an ongoing dispute on the interpretation of the "neutron puzzle" (for the latest arguments
see Ref.~\cite{Kuznetsov:2017qmo}).

\section{Ground state heavy baryons}
\label{sec:HB}

\begin{figure}[h]
\centering
\includegraphics[height=4cm]{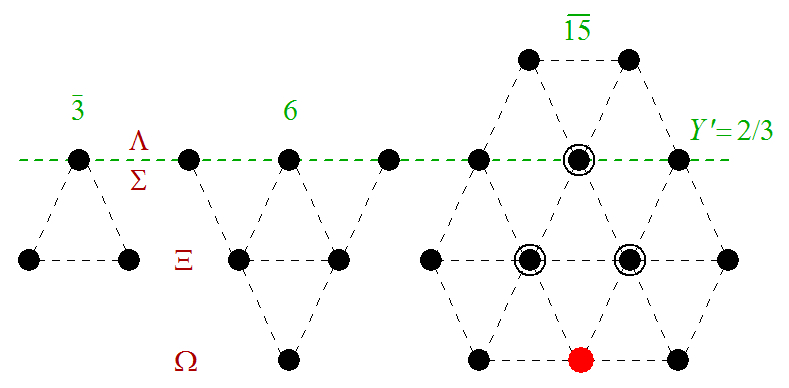}
\caption{Rotational
band of a soliton with one valence quark stripped off. Soliton spin
corresponds to the isospin $T^{\prime}$ of states on the quantization line
$Y^{\prime}=2/3$. We show three lowest allowed representations: antitriplet of
spin 0, sextet of spin 1 and the lowest exotic representation $\overline
{\mathbf{15}}$ of spin 1 or 0.  Heavy quark has to be added.}
\label{fig:irreps}%
\end{figure}

Recently wa have proposed~\cite{Yang:2016qdz}, following Ref.~\cite{Diakonov:2010zz}, 
to generalize the above approach to heavy baryons,
by stripping off one valence quark from the $K^P=0^+$ level, as shown in Fig.~\ref{fig:levels}.b,
and replacing it by a heavy quark to neutralize the color. 
In the large $N_c$ limit both systems: light and heavy baryons are described essentially by the same mean field, and the
only difference is now in the quantization condition:
\begin{itemize}
\item allowed SU(3) representations must contain states with hypercharge
$Y^{\prime}=(N_{c}-1)/3$.
\end{itemize}
The lowest allowed SU(3) representations are in this case (as in the quark model)
 $\overline{\mathbf{3}}$ of spin 0  and 
to ${\mathbf{6}}$ of spin 1 shown in Fig.~\ref{fig:irreps}. Therefore, the baryons constructed  from such a soliton and a heavy
quark form an SU(3) antitriplet of spin 1/2 and two sextets  of spin 1/2 and 3/2 that are subject to a hyper-fine
splitting. 

It is interesting to look at the 
 so called \emph{model-independent
  relations}\footnote{The term \textit{model-independent relations} in
  the present context has been first used in
  Ref.~\cite{Adkins:1984cf}.} 
   generated by the symmetries of the $\chi$QSM that include
   Gell-Mann--Okubo SU(3) mass relations within given baryon multiplet,
   and the so called Guadaginini~\cite{Guadagnini:1983uv} relations that follow from the {\em hedgehog}
   symmetry and relate mass splittings between different multiplets. 
The model predicts equal mass splittings separately in the
$\overline{\bm{3}}$ and $\bm{6}$, which are proportional to the hypercharge $Y$
\[
\Delta M=\delta_{\overline{\bm{3}}, \bm{6}}\, \times \, Y
\]
that are
independent of the spin and of the heavy quark mass. 
These relations
are indeed very well satisfied. For the $\bar{\bf 3}$ we have (in
MeV): 
\begin{align}
-\delta_{\overline{\bm 3}}
=\left. 182.9 \pm 0.3
\right|_{\Xi_c-\Lambda_c}
=\left. 173.6  \pm 0.7
\right|_{\Xi_b-\Lambda_b}, 
\label{eq:deltabar3}
\end{align}
which is satisfied with 7~\% accuracy. In the case of  the $\bm{6}$ we 
have more relations (in MeV): 
\begin{align}
 -\delta_{\bm{6}}& 
= \left. 123.3 \pm 2.1
  \right|_{\Xi^{\prime}_c-\Sigma_c}
= \left. 118.4 \pm 2.7
  \right|_{\Omega_c - \Xi^{\prime}_c} 
\cr
& 
= \left. 127.8 \pm 0.8
  \right|_{\Xi^{\ast}_c-\Sigma^{\ast}_c}
= \left. 120.0 \pm 2.0
  \right|_{\Omega^{\ast}_c - \Xi^{\ast}_c}
\cr
&
= \left. 121.6 \pm 1.3
  \right|_{\Xi^{\prime}_b-\Sigma_b}
= \left. 113.0 \pm 1.9
  \right|_{\Omega_b - \Xi^{\prime}_b}
\cr 
&
= \left. 121.7 \pm 1.3
  \right|_{\Xi^{\ast}_b-\Sigma^{\ast}_b}.
 \label{eq:delta6}
\end{align}
We see that the equality of splittings is quite accurate (at the 6~\%
level).  On the other hand the model {\em predicts} the values of the splitting parameters~\cite{Yang:2016qdz}
using as an input mass splittings of the light baryons:
\begin{align}
\delta_{\overline{\bm{3}}} = -203.8\pm3.5,\;\;\;\;
  \delta_{\bm{6}} = -135.2\pm3.3,
\label{eq:deltamodel}
\end{align}
in units of MeV, which overestimate the model-independent
determination (\ref{eq:deltabar3}, \ref{eq:delta6}) by approximately
13~\%. This is entirely within the expected accuracy of this approach.

In order to remove the degeneracy between sextet spin 1/2 and 3/2 multiplets, 
we introduce the spin-spin interaction hamiltonian expressed as:
\begin{align}
H_{LQ} = \frac{2}{3}\frac{\kappa}{m_{Q}\,M_{\mathrm{sol}}}\bm{J} 
\cdot  \bm{S}_{Q} 
= \frac{2}{3}\frac{\varkappa}{m_{Q}}
\bm{J} \cdot \bm{S}_{Q} 
\label{eq:ssinter}
\end{align}
where $\varkappa$ denotes the flavor-independent hyperfine coupling. The
operators ${\bm  J}$ and ${\bm S}_Q$ represent the spin
operators for the soliton and the heavy quark,
respectively. $M_{\mathrm{sol}}$ has been incorporated into an unknown
coefficient $\varkappa$. The Hamiltonian $H_{LQ}$ does not affect the
$\overline{\bm{3}}$ states, since  in this case $J=0$. In
${\bm{6}}$ $J=1$, and it couples to $\bm{S}_{Q}$
producing two multiplets $S=1/2$ and $S=3/2$.

We arrive therefore at another set of model-independent relations, which are not directly related to
the specifics of the soliton model, but provide a test of our
assumption concerning the spin interactions of
Eq.~(\ref{eq:ssinter}):
\begin{align}
\frac{\varkappa}{m_c}
&= \left. 64.5\pm0.8 \right|_{\Sigma_c}
 = \left. 69.1\pm2.1 \right|_{\Xi_c} 
 = \left. 70.7\pm2.6 \right|_{\Omega_c} 
\cr
\frac{\varkappa}{m_b}
&= \left. 20.2\pm1.9 \right|_{\Sigma_b}
 = \left. 20.3\pm0.1 \right|_{\Xi_b}.
\label{eq:spintest}
\end{align}
(in MeV). From the ratios of the spin splittings (\ref{eq:spintest}) we can determine
the ratio of the heavy-quark masses 
\begin{align}
\frac{m_c}{m_b}=0.29 - 0.31.
\label{eq:cbratio}
\end{align}
The experimental values of the $\overline{\rm MS}$ heavy quark masses  
lead to $m_c/m_b=0.305$  where both masses
$m_Q$ are evaluated at the renormalization point  $\mu=m_Q$~\cite{Patrignani:2016xqp}.

Apart from equal splittings in {\bf 6} (\ref{eq:delta6}), 
the model admits a sum rule relating particles from two spin
1/2 and spin 3/2 sextets (the latter denoted with an asterisk):
\begin{align}
M_{\Omega_{Q}^{\ast}} 
= 
 2M_{\Xi^{\prime}_{Q}}
 +M_{\Sigma_{Q}^{\ast}} 
 -2 M_{\Sigma_{Q}}.
\label{eq:OMpred}
\end{align}
Equation~(\ref{eq:OMpred}) yields $(2764.5\pm3.1)$~MeV for
$M_{\Omega_{c}^{\ast}}$, which is 1.4~MeV below the experiment, 
and predicts 
\begin{align}
M_{\Omega_{b}^{\ast}}=(6076.8\pm2.25)\;\mathrm{MeV}
\label{eq:Omstb2}
 \end{align}
for yet unmeasured $\Omega_{b}^{\ast}$.

\begin{figure}[h]
\centering
\includegraphics[height=7cm]{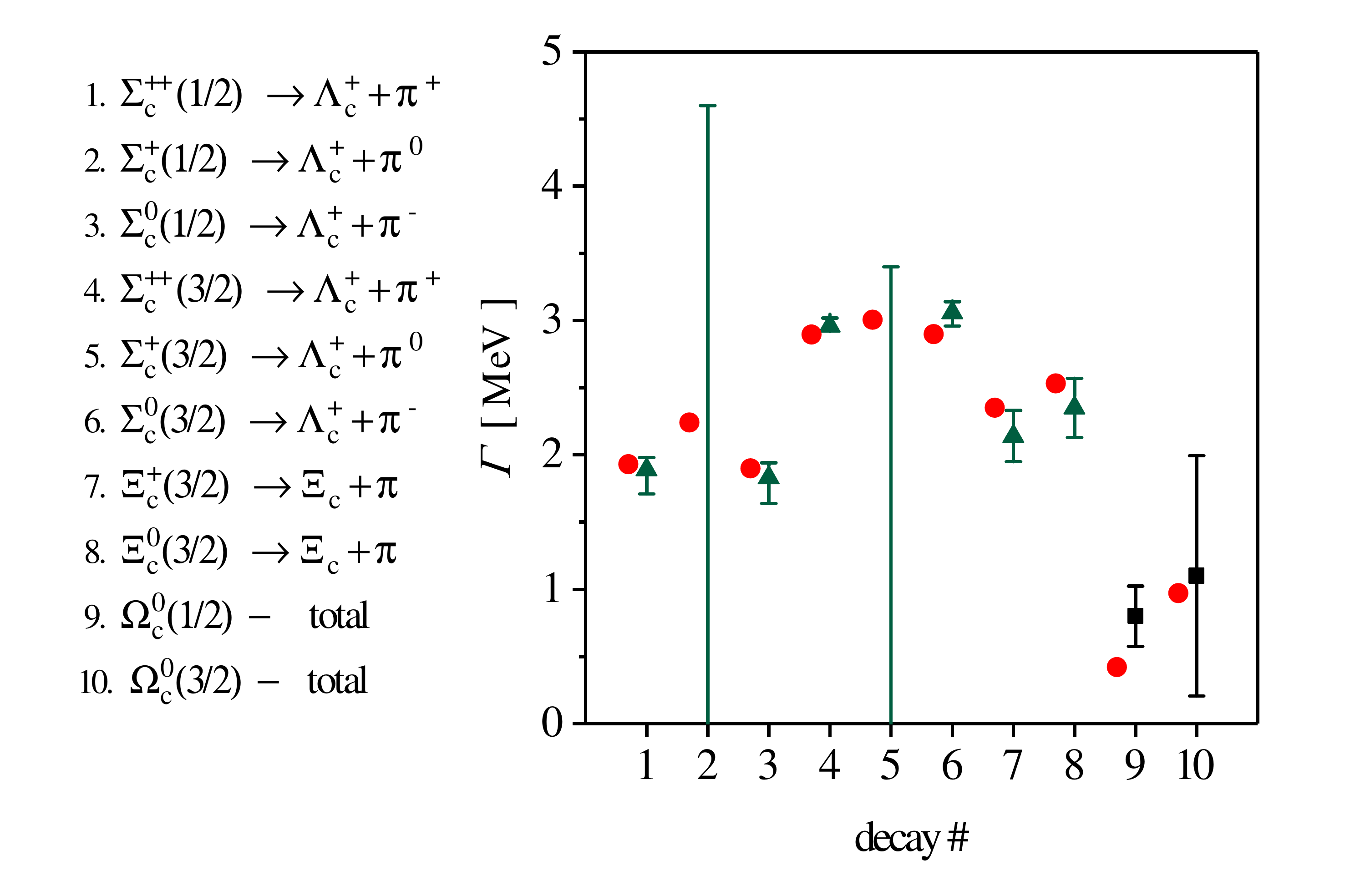}
\caption{Decay
widths of the charm baryons.  Red full circles correspond to our
theoretical predictions. Dark green triangles correspond to the experimental
data \protect{\cite{Patrignani:2016xqp}}. 
Data for decays 4 -- 6 of 
$\Sigma_{c}(\mathbf{6}_{1},3/2)$ have been divided by a factor
of 5 to fit within the plot area. Widths of two LHCb
\cite{Aaij:2017nav}
$\Omega^0_{c}$ states that we interpret as pentaquarks are plotted as black full
squares with  theoretical values shown as red full circles. }%
\label{fig:charmwidths}%
\end{figure}

Another phenomenological test of the model consists in the calculation of the decay widths.
Indeed, using heavy baryon chiral perturbation theory \cite{Cheng:2006dk} 
and the Goldberger-Treiman relation \cite{Yan:1992gz} one can derive
the formulae for the $p$-wave strong decay widths of the ground state sextets:
\begin{align}
\Gamma_{\Sigma(\boldsymbol{6}_{1})\rightarrow\Lambda(\overline{\boldsymbol{3}%
}_{0})+\pi} &  =\frac{1}{72\pi}\frac{p^{3}}{F_{\pi}^{2}}\frac{M_{\Lambda
(\overline{\boldsymbol{3}}_{0})}}{M_{\Sigma(\boldsymbol{6}_{1})}}%
H_{{\overline{\boldsymbol{3}}}}^{2}\frac{3}{8},\nonumber\\
\Gamma_{\Xi(\boldsymbol{6}_{1})\rightarrow\Xi(\overline{\boldsymbol{3}}%
_{0})+\pi} &  =\frac{1}{72\pi}\frac{p^{3}}{F_{\pi}^{2}}\frac{M_{\Xi
(\overline{\boldsymbol{3}}_{0})}}{M_{\Xi(\boldsymbol{6}_{1})}}H_{{\overline
{\boldsymbol{3}}}}^{2}\frac{9}{32}
\label{eq:widths1}%
\end{align}
where
\begin{align}
H_{{\overline{\boldsymbol{3}}}}=-a_{1}+\frac{1}{2}a_{2}.
\label{eq:H}
\end{align}
Constants $a_i$ enter the definition of the axial-vector
current~\footnote{
For the reader's convenience we give the relations of the constants
$a_{1,2,3}$ to nucleon axial charges in the chiral limit:  
$g_{A}=\frac{7}{30}\left(
  -a_{1}+\frac{1}{2}a_{2}+\frac{1}{14}a_{3}\right)$, $g_A^{(0)}=\frac
12 a_3$,  
$g_{A}^{(8)}=\frac{1}{10 \sqrt 3}\left(
  -a_{1}+\frac{1}{2}a_{2}+\frac{1}{2}a_{3}\right)$} 
and have been extracted from the
semileptonic decays of  the
  baryon octet in Ref.~\cite{Yang:2015era}. Model predictions (which are parameter free with one
  modification related to the rescaling of $a_1$ due to the fact that contrary to the light sector 
  we have $N_c-1$ valence quarks
  rather than $N_c$) for the partial decay widths
  are compared with the experimental data in Fig.~\ref{fig:charmwidths}. Surprisingly
  good agreement between predicted decay widths and the experimental data encourages us
  to apply the model to the excited heavy baryons.
  
\section{Excited heavy baryons }

Two possible kinds of excitations are present in the $\chi$QSM. Firstly, higher
SU(3) representations, similar to the antidecuplet in the light sector, appear
in the rotational band of the 
soliton of Fig.~\ref{fig:levels}.b.
The lowest possible exotic SU(3) representation is 
$\overline{\mathbf{15}}$ of positive parity  and spin 1
 ($\overline{\mathbf{15}}$ of spin 0 is heavier) 
depicted in Fig.~\ref{fig:irreps}. Second possibility corresponds to the
excitation of the sea quark from the $K^P=1^{-}$ sea level to the valence level~\cite{Diakonov:2010zz}
shown in Fig.~\ref{fig:levels}.c (or alternatively valence quark excitation to the first excited level~\footnote{We thank Victor
Petrov for pointing out this possibility.}
of $K^P=1^{-}$). In this case the parity is negative but the rotational band is the same 
as in Fig.~\ref{fig:irreps} with, however, different quantization condition:
\begin{itemize}
\item the isospin $\bm{T}^{\prime}$ of the states with $Y^{\prime}%
=(N_{c}-1)/3$ couples with the soliton spin $\bm{J}$ as follows:
$\bm{T}^{\prime}+\bm{J}=\bm{K}$, where $\bm{K}$ is the grand spin
of the excited level.
\end{itemize}

The first allowed SU(3) representation for one quark excited soliton is again
$\overline{\mathbf{3}}$ with $T^{\prime}=0$, which -- according to
the above condition for $K=1$  is quantized as spin $J=1$. We therefore
expect two hyper-fine split anti-triplets of spin 1/2 and 3/2.
For each of them
the $m_s$ splitting parameter $\delta_{\overline{\mathbf{3}}}^{\prime}$  is given by the same formula as for the ground state
antitriplet, and therefore we know its numerical value \cite{Yang:2016qdz}:
\begin{equation}
\delta_{\overline{\mathbf{3}}}^{\prime}=\delta_{\overline{\mathbf{3}}}=-180~\mathrm{MeV}.\label{eq:delta3bar}%
\end{equation}
There are rather well measured candidates for both anti-triplets:
for $(1/2)^{-}$ we have $\Lambda_{c}(2592)$ and $\Xi_{c}(2790)$ and for
$(3/2)^{-}$ there exist $\Lambda_{c}(2628)$ and $\Xi_{c}(2818)$. From this
assignment we get $\delta^{\prime}_{\overline{\mathbf{3}}}=-198$~MeV and
$-190$~MeV respectively, in relatively good agreement with
Eq.(\ref{eq:delta3bar}). Furthermore, we can extract hyper-fine splitting parameter:
\begin{align}
\frac{\kappa^{\prime}}{m_{c}} & = 
30~\mathrm{MeV}
\end{align}
which is different from Eq.~(\ref{eq:spintest}) since it corresponds to a different soliton configuration,
namely the one of Fig.~\ref{fig:levels}.c

\begin{figure}[h]
\centering
\includegraphics[width=5cm]{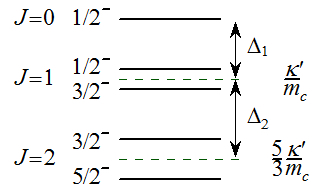} \vspace{-0.2cm}\caption{Schematic
spectrum of the $\Omega_Q$ members of the excited sextets.}%
\label{fig:spectrum6}%
\end{figure}

Next negative parity excitation is the SU(3) sextet (with $T'=1$, see Fig.~\ref{fig:irreps}), which for $K=1$
gives the soliton spin $J=0,1,2$ in close analogy to the total angular momentum of light subsystem in the
quark model.  By adding one heavy quark we end up with five possible total spin $S$ excitations: 
for $J=0$: $S=1/2$, for $J=1$: $S=1/2$ and 3/2, 
and for $J=2$:  $S=3/2$ and 5/2. The $m_s$ mass splittings depend in this case on $J$ and read
\begin{equation}
\delta_{\mathbf{6}\,J}^{\prime} =\delta_{\mathbf{6}}-\frac{3}{20}\delta
\times \left\{
\begin{array}
[c]{rcc}%
2 & \text{for} & J=0\\
1 & \text{for} & J=1\\
-1 & \text{for} & J=2
\end{array}
\right.  ,
\end{equation}
where $\delta_{\mathbf{6}}=-120$~MeV \cite{Yang:2016qdz}
corresponds to the ground state sextet
splitting. Unfortunately we do not know the value of a new parameter
$\delta$ corresponding to the quark transition depicted in Fig.~\ref{fig:levels}.c,
which is absent in the case of light baryons.
It is however possible to eliminate the unknown parameters (not only $\delta$
but also another parameter in the rotational hamiltonian) and obtain spectrum 
of the $\Omega_Q$ members of the excited sextets depicted
in Fig.~\ref{fig:spectrum6}. We have checked in Ref.~\cite{Kim:2017jpx} that it is 
impossible to fit all five LHCb states into the pattern of Fig.~\ref{fig:spectrum6}.

\begin{figure}[h]
\centering
\includegraphics[width=9.0cm]{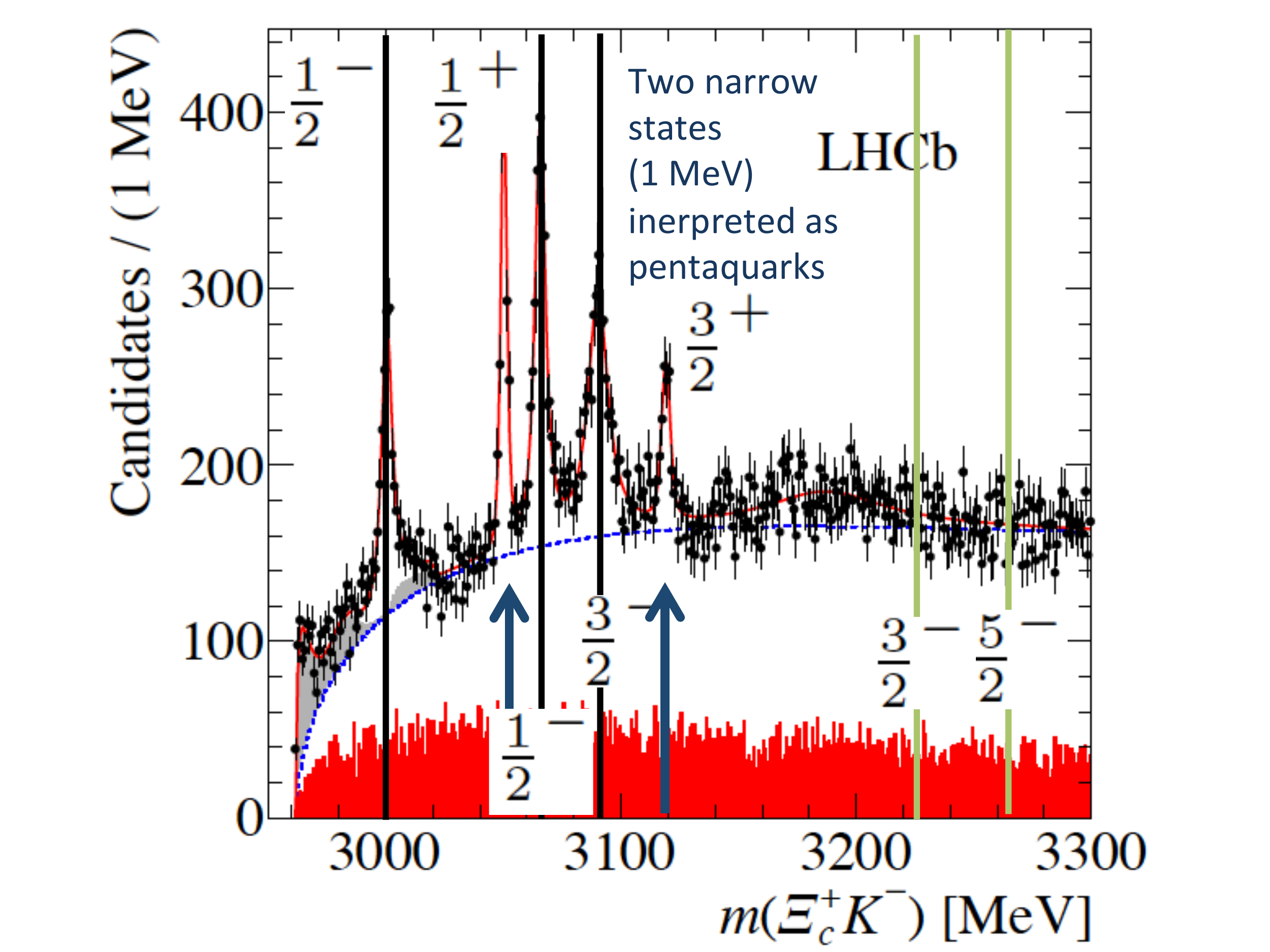} \vspace{-0.2cm}
\caption{Spectrum of the $\Omega^0_c$ states (from Ref.[8]) with theoretical predictions of the present model}%
\label{fig:Omegas}%
\end{figure}

We have
therefore {\em forced} model constraints of Fig.~\ref{fig:spectrum6},
which allows to accommodate only three out of five LHCb states (see black vertical lines in Fig.~\ref{fig:Omegas}). 
Two heaviest $\chi$QSM states (green lines in  Fig.~\ref{fig:Omegas}) lie already above the decay threshold to heavy mesons,
and it is quite possible that they have very small branching ratio to the  $\Xi^+_c+K^-$ final state analyzed by the LHCb. Two remaining states
indicated by dark-blue arrows in Fig.~\ref{fig:Omegas}, which are 
hyper fine  split by 70~MeV  (as the ground state sextets that belong to the same rotational band, see the first line of Eq.~(\ref{eq:spintest})), 
can be therefore interpreted as the members of exotic 
$\overline{\mathbf{15}}$ of positive parity shown as a red dot in Fig.~\ref{fig:irreps}. This interpretation is reinforced
by the decay widths, which can be computed in the model:
\begin{align}
\Gamma_{\Omega(\overline{\boldsymbol{15}}_{1})\rightarrow\Xi(\overline
{\boldsymbol{3}}_{0})+K} &  =\frac{1}{72\pi}\frac{p^{3}}{F_{K}^{2}}%
\frac{M_{\Xi(\overline{\boldsymbol{3}}_{0})}}{M_{\Omega(\overline
{\boldsymbol{15}}_{1})}}G_{{\overline{\boldsymbol{3}}}}^{2}\frac{3}%
{10},\nonumber\\
\Gamma_{\Omega(\overline{\boldsymbol{15}}_{1})\rightarrow\Omega(\boldsymbol{6}%
_{1})+\pi} &  =\frac{1}{72\pi}\frac{p^{3}}{F_{\pi}^{2}}\frac{M_{\Omega
(\boldsymbol{6}_{1})}}{M_{\Omega(\overline{\boldsymbol{15}}_{1})}%
}G_{\boldsymbol{6}}^{2}\frac{4}{15}\gamma,\nonumber\\
\Gamma_{\Omega(\overline{\boldsymbol{15}}_{1})\rightarrow\Xi(\boldsymbol{6}%
_{1})+K} &  =\frac{1}{72\pi}\frac{p^{3}}{F_{K}^{2}}\frac{M_{\Xi(\boldsymbol{6}%
_{1})}}{M_{\Omega(\overline{\boldsymbol{15}}_{1})}}G_{\boldsymbol{6}}^{2}%
\frac{2}{15}\gamma
\label{eq:Omegawidths}%
\end{align}
where coefficients $\gamma$ depend on the initial and final spin of the involved baryons:
\begin{align}
\gamma(1/2\rightarrow1/2)   ={2}/{3}, \qquad &
\gamma(1/2\rightarrow3/2)   ={1}/{3},\nonumber\\
\gamma(3/2\rightarrow1/2)   ={1}/{6}, \qquad &
\gamma(3/2\rightarrow3/2)   ={5}/{6}.
\end{align}
Here \begin{align}
 G_{\overline{\boldsymbol{3}}}&=-a_{1}-\frac{1}{2}a_{2}, \notag \\
G_{\boldsymbol{6}}&=-a_{1}-\frac{1}{2}a_{2}-a_{3}.
\label{eq:Gs}
\end{align}
These widths are of the order of 1~MeV and agree with the
LHCb measurement (see Fig.~\ref{fig:charmwidths}). Such small widths are in fact expected in the present approach,
since the leading $N_c$ terms of the  couplings $G_{\overline{\boldsymbol{3}}}$ and $G_{\boldsymbol{6}}$ cancel 
in the non-relativistic limit.

\section{Summary}

In the large $N_c$ limit both heavy and light baryons are described by the universal mean-field. This allows us
to relate the properties of heavy baryons to the light ones. We have shown in Ref.~\cite{Yang:2016qdz} that the universal mean field
gives simultaneously good description of the ground-state
$\overline{\mathbf 3}$ and ${\mathbf 6}$ multiplets of heavy 
baryons.   In Ref.~\cite{Kim:2017jpx}
 we have demonstrated that the same picture predicts the following
excited states for heavy-quark baryons:
\begin{itemize}
\item two hyper-fine split ($1/2^-$ and $3/2^-$)  $\overline{\mathbf 3}'$
which experimentally have very good candidates,
\item five excited sexstets (rotationally and hyper-fine split) with
quantum numbers $(J=0,1/2^-)$, $(J=1,1/2^-,3/2^-)$ and
$(J=2,3/2^-,5/2^-)$, where $J$ denotes the soliton spin, 
\item two hyper-fine split exotic $\overline{\mathbf {15}}$-plets
  with quantum numbers $1/2^+$ and $3/2^+$. 
\end{itemize}

 The observation of the new excited $\Omega^0_c$'s allows us to get
 insight into the excited sextets and $\overline {\mathbf{15}}$-plets. 
We identify the observed $\Omega_c(3000)$, $\Omega_c(3066)$ and
$\Omega_c(3090)$ with $\left(J=0:~ 1/2^-\right)$ and $\left(J=1:~
  1/2^-,3/2^-\right)$ states from the excited sextet, whereas 
   the most narrow $\Omega_c(3050)$ and $\Omega_c(3119)$ states
   we
identify
with  $\left(J=1:~ 1/2^+,3/2^+\right)$ states from {the} exotic 
$\overline{\mathbf {15}}$ multiplet. The remaining two $\left(J=2:~
  3/2^-,5/2^-\right)$ states from the sextet have masses above
the $\Xi+D$ threshold (3185~MeV), so they are probably hidden in a large
bump observed by the LHCb collaboration above 3200~MeV. It should be
stressed that the simplest scenario in which all five LHCb $\Omega_c^0$
states are classified as members of the excited sextets contradicts
general mass formulae derived within the $\chi$QSM.

The simplest way to confirm or falsify our identification is to search 
 for the {\it isospin} partners of $\Omega^0_c$ from the $\overline{\mathbf
   {15}}$. For example, they can be searched in the mass distribution of
 $\Xi_c^0+K^-$ or $\Xi_c^+ + \bar K^0$, the $\Omega^0_c$'s from the
 sextet do not decay into these channels. 
 
 \section*{Acknowledgments}
It is a pleasure to thank  H.-C.~Kim, M.V.~Polyakov and G.S.~Yang 
for a fruitful collaboration that led to the research reported in this paper.
I would like to thank
the organizers of the Corfu Summer Institute 2017 'School and Workshops on Elementary Particle Physics and Gravity'
where these results were presented,
for very stimulating and interesting meeting.

\end{document}